\documentclass[twocolumn]{aastex61}

\newcommand{\obj}{A/2017~U1}
\newcommand{\kms}{km~sec$^{-1}$}
\newcommand{\mearth}{M$_{\oplus}$}
\newcommand{\msun}{M$_{\odot}$}

\submitjournal{Research Notes of the AAS}

\begin{document}

\title{Origin of Interstellar Object \obj{} in a Nearby Young Stellar Association?}

\correspondingauthor{Eric Gaidos}
\email{gaidos@hawaii.edu}

\author[0000-0002-5258-6846]{Eric Gaidos}
\affil{Department of Geology \& Geophysics, University of Hawai`i at M\={a}noa, Honolulu, HI 96822}

\author{Jonathan P. Williams}
\affil{Institute for Astronomy, University of Hawai`i at M\={a}noa, Honolulu, HI 96822}

\author{Adam Kraus}
\affil{Department of Astronomy, University of Texas at Austin, Austin, TX 78712}

\section*{}
\obj{} is a minor body on a hyperbolic orbit that is presumably of interstellar origin.  It exhibited no coma after perihelion (K. Meech, MPEC 2017-U183) and a low signal-to-noise spectrum lacks features and is redder than any known asteroid group \citep{Masiero2017}.  These observations suggest absence of ices, as its surface reached 550~K (assuming 5\% albedo) during perihelion, and an origin close to another star.  

The pre-encounter motion of \obj{} is another clue to its origin.  From the JPL HORIZONS solution its velocity ($+U$ towards the Galactic center) was $UVW = (-11.29 \pm 0.06$,$-22.36\pm0.06$, $-7.61\pm0.07)$ \kms{}.  The radial velocity and impact parameter are RV$_0 = -26.18\pm 0.09$ \kms{}, $b = 0.849 \pm 0.003$~AU.  

We searched the Tycho-Gaia Astrometric Solution catalog \citep{Gaia2016} for stars with matching space motions, selecting candidates with proper motions that lie along (but opposite in direction to) the great circle between star and \obj{} radiant, estimating the best-fit RVs, and comparing those with $\chi^2 < 7.8$ to literature values.  HR~4834, HD~86427, HD~137957, HD~140241 have observed-predicted $UVW$ differences $\lesssim 3$~\kms{}, three of these are putative members of the 10-15~Myr Scorpius-Centaurus association \citep[For another search amoung nearby stars see][]{Mamajek2017}.  Indeed, \obj{}, moves within $\lesssim 10$~\kms of the Local Standard of Rest \citep{Coskunoglu2011} and thus many star forming regions and young stars.  The difference w.r.t. the LSR is almost entirely in $V$, the component of LSR motion with the largest discrepancies between estimates \citep{Robin2017}.  Over time, passing stars and clouds would have perturbed \obj{} away its initial LSR-like motion, thus the deviation in $U$ and $W$ of only a few \kms{} points to an age $\ll 1$~Gyr \citep{Robin2017}, or coincidence.  

This suggests a possible origin in a nearby young stellar cluster or association.  The best kinematic matches ($\lesssim 2$ \kms, within uncertanties) are to the Carina and Columba Associations, which could be part of a larger star-forming event with the Tucana-Horologium associations \citep{Torres2008}.  Estimated distances are 50-85~pc \citep{Torres2008,Malo2014} and isochrone ages are $\approx45$~Myr \citet{Bell2015}, thus an object ejected at 1-2~\kms{} soon after star fromation could travel to the present location of \obj{} (and the Sun) by the present time.

\begin{figure*}
\plottwo{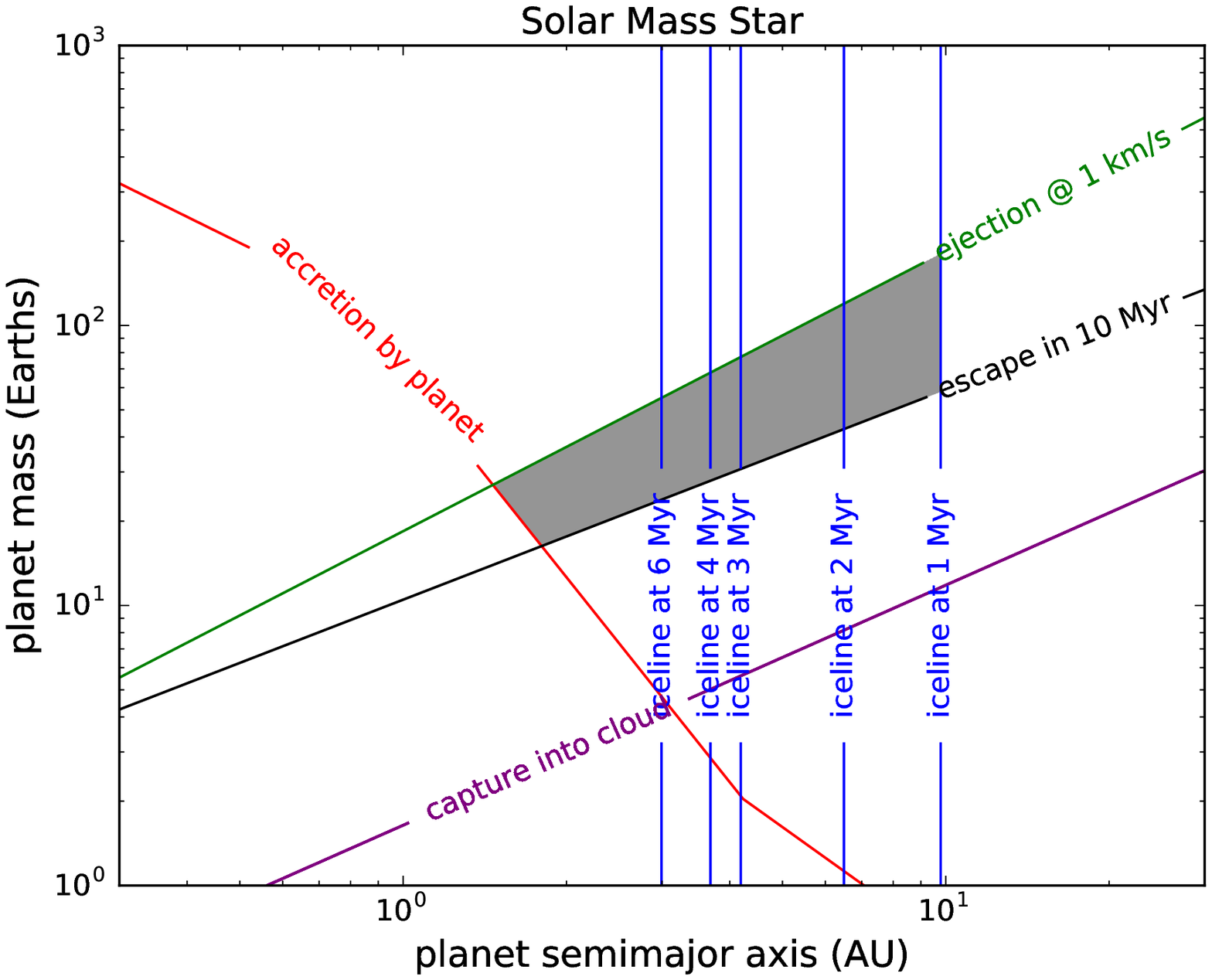}{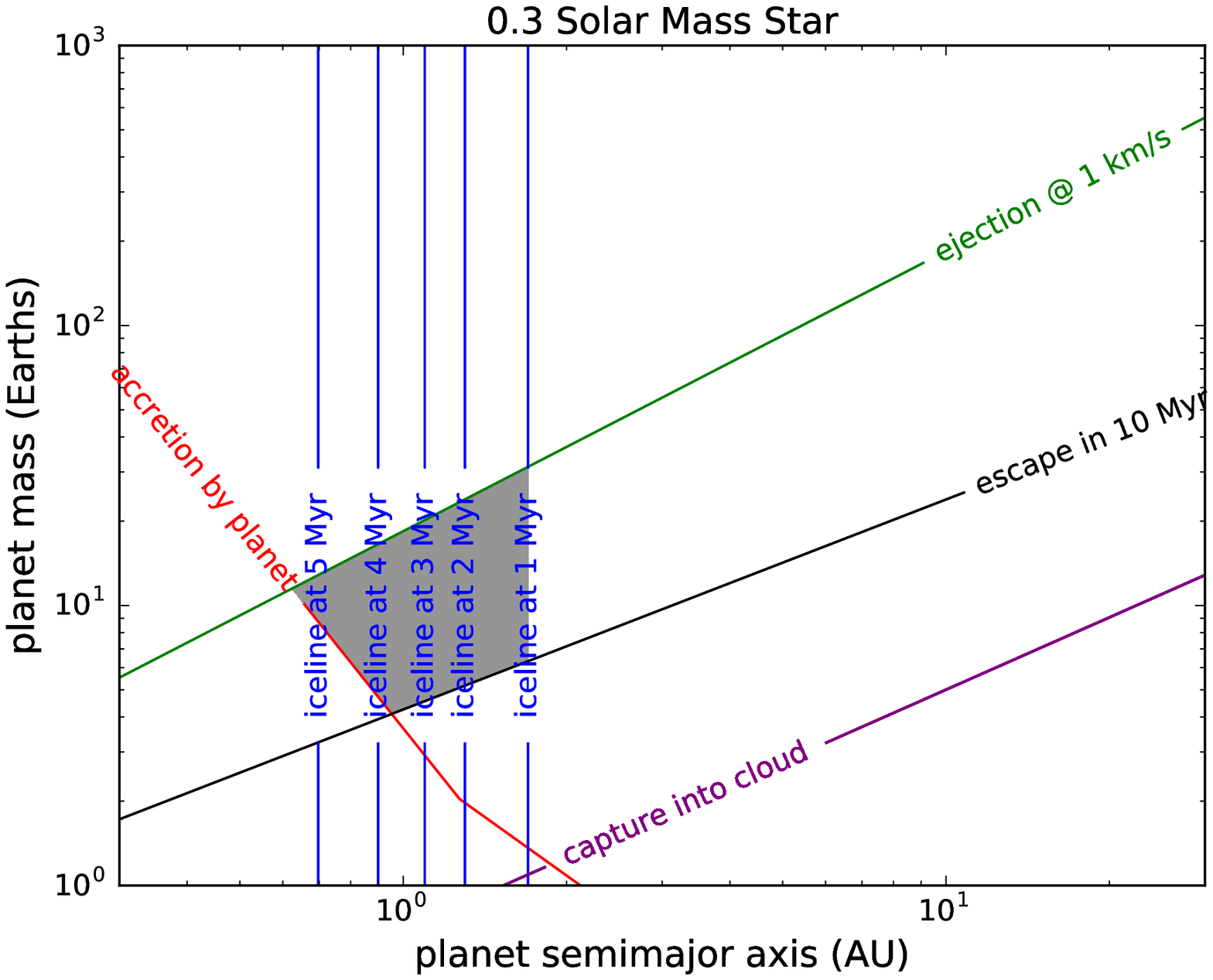}
\caption{Allowed mass and semimajor axis of the hypothetical solitary planet that ejected \obj{} from a protoplanetary disk around 1 or 0.3 \msun{} star.  Below the red line planets accrete rather than scatter planetesimals.  Above the green line planets eject planetesimals at $>1$~\kms{}.  Below the purple line planetesimals are captured into clouds by the cluster tide.  Below the black line planetesimals require $>10$~Myr to escape.  To the right of the blue lines planetsimals contain ices. \label{fig.one}}
\end{figure*}

We suggest that \obj{} formed in a protoplanetary disk in the Carina/Columba associations and was ejected by a planet $\approx$40~Myr ago. The absence of ice indicates an origin inside the ``ice line" of the disk plus an ejection velocity of 1-2~\kms{} \citep[assuming the cluster was already unbound,][]{Williams2007}, constrain the mass $m_P$  and semi-major axis $a_P$ of the planet.   Figure \ref{fig.one} shows the allowed ranges of $a_P$ and $m_p$ for 1 and 0.3\msun{} stars, bounded by: (1) the planet scatter rather than accrete the planetesimal \citep[red line, using the planet mass-radius relation of ][]{Chen2017}; (2) the escape energy, equated to a single step of a randomin total energy and scaling as $Gm_P/a_p$ \citep{Tremaine1993,Gaidos1995}, is $\lesssim 1$~\kms{} (green line); (3) the object is not captured into a compact exo-Oort cloud due to cluster tides \citep[purple line,]{Gaidos1995}; (4) the time to escape the star is $<10$~Myr (i.e., $\ll$ than the association age); and (5) $a_p$ is inside the iceline before ejection \citep{Zhang2015}.  Permitted values (grey zone) center around a 20-30\mearth{} planet forming by a few Myr within a few AU, reminiscient of the core accretion scenario for giant planet formation.  In contrast, a ``super-Earth'' at $\sim$1~AU could eject ice-free planetesimals from a lower-mass M dwarf (Fig. \ref{fig.one}b).  A lower limit on the space density of such objects (using $b$, RV$_0$ and one detection in 7~years of PAN-STARRS operation) is $1 \times 10^{14}\,{\rm pc}^{-3}$.  Adopting a 230~m diameter based on $H=22.08$, the mass density is $\sim 0.2$\mearth{}~pc$^{-3}$; $\sim 1$\mearth{} per extant star, or a few \% of the total solids in each protoplanetary disk.  We predict that future interlopers could radiants similar to \obj{}: Watch this space.    
 
\end{document}